\documentclass[twocolumn,prl]{revtex4-1}
\usepackage{graphicx,color}
\usepackage{bm}

\begin{document}

\title{
Role of the spin-orbit coupling in the Kugel-Khomskii model
on the honeycomb lattice
}

\author{Akihisa Koga}
\affiliation{
  Department of Physics, Tokyo Institute of Technology,
  Meguro, Tokyo 152-8551, Japan
}

\author{Shiryu Nakauchi}
\affiliation{
  Department of Physics, Tokyo Institute of Technology,
  Meguro, Tokyo 152-8551, Japan
}

\author{Joji Nasu}
\affiliation{
  Department of Physics, Tokyo Institute of Technology,
  Meguro, Tokyo 152-8551, Japan
}

\date{\today}

\begin{abstract}
We study the effective spin-orbital model for honeycomb-layered transition metal compounds,
applying the second-order perturbation theory
to the three-orbital Hubbard model with the anisotropic hoppings.
This model is reduced to the Kitaev model
in the strong spin-orbit coupling limit.
Combining the cluster mean-field approximations with
the exact diagonalization,
we treat the Kugel-Khomskii type superexchange interaction and
spin-orbit coupling on an equal footing to discuss ground-state properties.
We find that a zigzag ordered state is realized in the model within nearest-neighbor interactions.
We clarify how the ordered state competes with
the nonmagnetic state, which is adiabatically connected to
the quantum spin liquid state realized in a strong spin-orbit coupling limit.
Thermodynamic properties are also addressed.
The present work should provide another route
to account for the Kitaev-based magnetic properties in candidate materials.
\end{abstract}


\maketitle


Orbital degrees of freedom have been studied as a central topic of
strongly correlated electron systems as they possess own quantum dynamics
and are strongly entangled with other degrees of freedom such as charge
and spin~\cite{Tokura2000}.
Recently, multiorbital systems with strong spin-orbit (SO) couplings
have attracted considerable attention~\cite{Witczak2014,Watanabe}.
One of the intriguing examples is the series of the Mott insulators
with honeycomb-based structures such as
$A_2\rm IrO_3$ $(A=\rm Na, Li)$~\cite{Iridate1,Iridate2,Iridate3}, and
$\beta$-$\rm Li_2IrO_3$~\cite{LiIrO}.
In these compounds, a strong SO coupling for $5d$ electrons
lifts the triply degenerate $t_{2g}$ levels and
the low-energy Kramers doublet, which is referred to as an isospin, plays
an important role at low temperatures.
Furthermore, anisotropic electronic clouds intrinsic
in the $t_{2g}$ orbitals result in peculiar exchange couplings and
the system is well described by the Kitaev model
for the isospins~\cite{Kitaev,Jackeli}.
The ground state of this model is a quantum spin liquid (QSL),
and hence a lot of experimental and theoretical works have been devoted
to the iridium oxides in this context~\cite{Katukuri,Yamaji,Nasu,Nasu2015,Yoshitake2016,Knolle2014,Knolle2014b,Suzuki,zigzag}.
Very recently, the ruthenium compound $\alpha$-$\rm RuCl_3$
with $4d$ electrons has been studied actively as another Kitaev candidate
material~\cite{RuCl,RuCl-Plumb14,RuCl-Sears15,RuCl-Majumder15,RuCl-Sandilands15,RuCl-Johnson15,RuCl-Koitzsch16,RuCl-Sandilands16,NasuNature}.
In general, the SO coupling in $4d$ orbitals is weaker than that
in $5d$ orbitals and is comparable with the exchange energy.
Therefore, it is highly desired to deal with SO and exchange couplings
on an equal footing
although the magnetic properties for honeycomb-layered compounds
have been mainly discussed 
within the isospin model with the Kitaev and other exchange couplings
including longer-range interactions
~\cite{zigzag,RuCl-Kim15,RuCl-Banerjee16,RuCl-Winter17,YamajiTPQ,RuCl-Do}.

In this Letter, we study the role of the SO coupling in the Mott insulator
with orbital degrees of freedom.
We examine the localized spin-orbital model with the Kugel-Khomskii type
superexchange interactions between nearest-neighbor sites and
onsite SO couplings on the two-dimensional honeycomb lattice.
In the strong SO coupling limit,
this model is reduced to the Kitaev model and
the QSL state is realized.
On the other hand, a conventional spin-orbital ordered state may be stabilized
in the small SO coupling case.
To examine the competition between the magnetically disordered and
ordered states in the intermediate SO coupling region,
we first use the cluster mean-field (CMF) theory~\cite{CMF}
with the exact diagonalization (ED).
We determine the ground state phase diagram in the model
and clarify that a zigzag magnetically ordered state is realized due to
the competition between distinct exchanges.
Calculating the specific heat and entropy in terms of
the thermal pure quantum (TPQ) state~\cite{TPQ},
we discuss how thermodynamic properties characteristic of the Kitaev model
appear in the intermediate SO coupling region.

\begin{figure}
\begin{center}
\includegraphics[width=7cm]{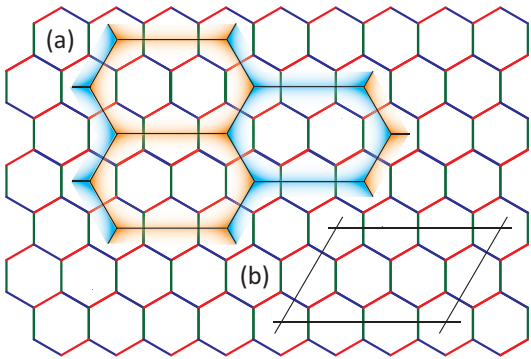}
\end{center}
\caption{
Honeycomb lattice.
(a) Effective cluster model with ten sites, which are treated
in the framework of the CMF method.
(b) Twelve-site cluster for the TPQ states.
}
\label{fig:lattice}
\end{figure}

We start with the three-orbital Hubbard model on the honeycomb lattice.
This should be appropriate to describe the electronic state of
the $t_{2g}$ orbitals in the compounds $A_2\rm IrO_3$ and
$\alpha$-$\rm RuCl_3$ since there exists a
large crystalline electric field for the $d$ orbitals.
The transfer integral $t$ between the $t_{2g}$ orbitals
via ligand $p$ orbitals are evaluated from the Slater-Koster parameters,
where the neighboring octahedra consisting of
six ligands surrounding transition metal ions share their edges.
Note that the transfer integrals involving one of the three $t_{2g}$ orbitals
vanish due to the anisotropic electronic clouds~\cite{Jackeli}.
We refer to this as an inactive orbital and the other orbitals as active ones.
These depend on three inequivalent bonds,
which are schematically shown as the distinct colored lines
in Fig.~\ref{fig:lattice}.
Moreover, we consider the onsite intra- and inter-orbital Coulomb interactions,
$U$ and $U'$, Hund coupling $K$, and pair hopping $K'$
in the conventional manner.
In the following, we restrict our discussions to the conditions
$U=U'+2K$ and $K'=K$,
which are lead by the symmetry argument of the degenerate orbitals.

We use the second-order perturbation theory
in the strong coupling limit
since the Mott insulating state is realized in the honeycomb-layered compounds.
We then obtain the Kugel-Khomskii-type exchange model,
assuming that five electrons occupy the $t_{2g}$ orbitals in each site.
By taking the SO coupling into account,
the effective Hamiltonian is explicitly given as
\begin{eqnarray}
{\cal H}=\sum_{\langle ij\rangle_\gamma}{\cal H}_{ij}^{{\rm ex}(\gamma)}-\lambda \sum_i {\bf L}_i\cdot{\bf S}_i,
\label{H}
\end{eqnarray}
where $\lambda$ is the SO coupling, and
${\bf S}_i$ and ${\bf L}_i$ are spin and orbital angular-momentum operators
at the $i$th site, respectively.
The exchange Hamiltonian ${\cal H}_{ij}^{{\rm ex}(\gamma)}$, which depends on the bond $\gamma(=x,y,z)$ of the honeycomb lattice (see Fig.~\ref{fig:lattice}), is given as
\begin{eqnarray}
{\cal H}^{{\rm ex}(\gamma)}_{ij}={\cal H}_{1;ij}^{(\gamma)}+{\cal H}_{2;ij}^{(\gamma)}+{\cal H}_{2;ij}^{(\gamma)},
\end{eqnarray}
with
\begin{widetext}
\begin{eqnarray}
{\cal H}_{1;ij}^{(\gamma)}&=&
2J_1\left({\bf S}_i\cdot{\bf S}_j+\frac{3}{4}\right)
\left[
\tau_{ix}^{(\gamma)}\tau_{jx}^{(\gamma)}-\tau_{iy}^{(\gamma)}\tau_{jy}^{(\gamma)}-\tau_{iz}^{(\gamma)}\tau_{jz}^{(\gamma)}
+\frac{1}{4}\tau_{i0}^{(\gamma)}\tau_{j0}^{(\gamma)}-\frac{1}{4}\left(\tau_{i0}^{(\gamma)}+\tau_{j0}^{(\gamma)}\right)
\right],\\
{\cal H}_{2;ij}^{(\gamma)}&=&2J_2\left({\bf S}_i\cdot{\bf S}_j-\frac{1}{4}\right)
\left[
\tau_{ix}^{(\gamma)}\tau_{jx}^{(\gamma)}-\tau_{iy}^{(\gamma)}\tau_{jy}^{(\gamma)}-\tau_{iz}^{(\gamma)}\tau_{jz}^{(\gamma)}
+\frac{1}{4}\tau_{i0}^{(\gamma)}\tau_{j0}^{(\gamma)}+\frac{1}{4}\left(\tau_{i0}^{(\gamma)}+\tau_{j0}^{(\gamma)}\right)
\right],\\
{\cal H}_{3;ij}^{(\gamma)}&=&-\frac{4}{3}(J_2-J_3)\left({\bf S}_i\cdot{\bf S}_j-\frac{1}{4}\right)
\left[
\tau_{ix}^{(\gamma)}\tau_{jx}^{(\gamma)}+\tau_{iy}^{(\gamma)}\tau_{jy}^{(\gamma)}-\tau_{iz}^{(\gamma)}\tau_{jz}^{(\gamma)}
+\frac{1}{4}\tau_{i0}^{(\gamma)}\tau_{j0}^{(\gamma)}
\right],\label{Hex}
\end{eqnarray}
\end{widetext}
where we follow the notation of Ref.~\cite{Prog}, and $J_1=2t^2/U [1-3K/U]^{-1}, J_2=2t^2/U [1-K/U]^{-1},
J_3=2t^2/U [1+2K/U]^{-1}$ are the exchange couplings between nearest neighbor spins.
Here, we have newly introduced the orbital pseudospin operators $\tau_l^{(\gamma)}$ with $l=x,y,z,0$.
Note that its definition depends on the direction of the bond ($\gamma$-bond) between the nearest neighbor pair $\langle ij\rangle$.
$\tau_l^{(\gamma)}$ is represented by the $3\times 3$ matrix
based on the three orbitals:
the $2\times 2$ submatrix on the two active orbitals is given by
$\sigma_l/2$ for $l=x,y,z$ and the identity matrix for $l=0$,
and the other components for one inactive orbital are zero,
where $\sigma_l$ is the Pauli matrix.
We here note that Hamiltonian ${\cal H}_1$ enhances ferromagnetic correlations,
while ${\cal H}_2$ and ${\cal H}_3$ lead to antiferromagnetic correlations.
Therefore, spin frustration should play an important role
for the ground state in the small $K/U$ region, where $J_1\sim J_2 \sim J_3$.

What is the most distinct from ordinary spin-orbital models is that
the present system describes not only
spin-orbital orders but also the QSL state realized in the Kitaev model.
When the SO coupling is absent,
the system is reduced to the standard Kugel-Khomskii type
Hamiltonian.
In the large Hund coupling case,
the Hamiltonian ${\cal H}_{1;ij}^{(\gamma)}$ is dominant.
Then, the ferromagnetically ordered ground state should be realized
despite the presence of orbital frustration.
In the smaller case of the Hund coupling,
the ground state is not trivial due to the existence of spin frustration,
discussed above.
On the other hand, in the case $\lambda\rightarrow \infty$,
the SO coupling lifts the degeneracy at each site and
the lowest Kramers doublet,
$|\tilde{\sigma}\rangle=(
|xy,\sigma\rangle\mp|yz,\bar{\sigma}\rangle
+i|zx,\bar{\sigma}\rangle)/\sqrt{3}$,
plays a crucial role for low temperature properties.
Then, the model Hamiltonian Eq.~(\ref{H}) is reduced
to the exactly solvable Kitaev model with the spin-1/2 isospin operator
$\tilde{\bf S}$,
as ${\cal H}_{\rm eff}=-\tilde{J}\sum_{\langle ij\rangle_\gamma}\tilde{S}_{i\gamma}\tilde{S}_{j\gamma}\; (\gamma=x,y,z)$,
where $\tilde{J}[=2(J_1-J_2)/3]$
is the effective exchange coupling~\cite{Kitaev}.
It is known that, in this effective spin model,
the QSL ground state is realized with the spin gap.
At finite temperatures, a fermionic fractionalization appears
together with double peaks in the specific heat~\cite{Nasu,Nasu2015}.
In the following, we set the exchange coupling $J_1$ as a unit of energy.
We then study ground-state and finite-temperature properties
in the spin-orbital system with parameters $K/U$ and $\lambda/J_1$.

First, we discuss ground state properties
in the spin-orbital model by means of the CMF method~\cite{CMF}.
In the method, the original lattice model is mapped to an effective
cluster model, where spin and orbital correlations in the cluster can be
taken into account properly.
Intercluster correlations are treated through several mean-fields
at $i$th site,
$\langle S_{ik}\rangle, \langle \tau_{il}^{(\gamma)}\rangle$
 and $\langle S_{ik}\tau_{il}^{(\gamma)}\rangle$,
where $k=x,y,z$ and $l=x,y,z,0$.
These mean-fields are determined via the self-consistent conditions
imposed on the effective cluster problem.
The method is comparable with the numerically exact methods
if the cluster size is large, and has successfully been applied to quantum
spin~\cite{CMF,SakaiQS,KawaguchiQS,YamamotoQS} and
hard-core bosonic systems~\cite{YamamotoB,HassanB,SuzukiB}.
To describe some possible ordered states
such as the zigzag and stripy states~\cite{zigzag}, 
we introduce two kinds of clusters in the honeycomb lattice,
which are shown as distinct colors in Fig.~\ref{fig:lattice}(a).
Using the ED method,
we self-consistently solve two effective cluster problems.
To discuss magnetic properties at zero temperature, we calculate
spin and orbital moments,
$m^\alpha_S=|\sum_i(-1)^{\delta^\alpha_i}\langle{\bf S}_i\rangle|/N$ and
$m^\alpha_L=|\sum_i(-1)^{\delta^\alpha_i}\langle{\bf L}_i\rangle|/N$,
where $N$ is the number of sites and $\delta^\alpha_i$
is the phase factor for an ordered state $\alpha$.

\begin{figure}
\begin{center}
\includegraphics[width=6cm,angle=-90]{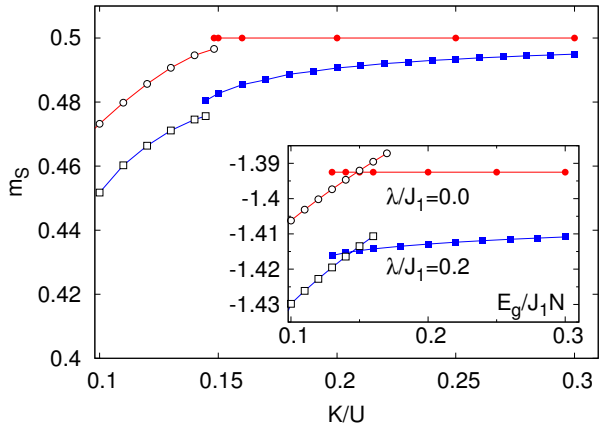}
\end{center}
\caption{
The spin moments as a function of the Hund coupling $K/U$.
Solid and open circles (squares) represent the results
for the ferromagnetically and zigzag ordered states
in the system with $\lambda/J_1=0.0$ $(0.2)$.
The ground state energy is shown in the inset.
}
\label{fig:mag}
\end{figure}

When $\lambda=0$, the spin and orbital degrees of freedom are decoupled.
Here, we show in Fig.~\ref{fig:mag} the spin moments
$m^f_S$ and $m^z_S$ for
the ferromagnetically and zigzag ordered states, respectively,
which are obtained by means of the ten-site CMF method (CMF-10).
Namely, we have confirmed that other ordered states
such as antiferromagnetic and stripy states are never stabilized
in the present calculations,
and thereby we do not show them in Fig.~\ref{fig:mag}.
Meanwhile, the local orbital moment disappears in the case $\lambda=0$.
In the system with the large Hund coupling,
the exchange coupling $J_1$ is dominant, and
the ferromagnetically ordered ground state is realized
with the fully-polarized moment $m^f_S=0.5$, as shown in Fig. \ref{fig:mag}.
On the other hand, in the smaller $K$ region,
the exchange couplings $J_2$ and $J_3$ are comparable with $J_1$.
Since ${\cal H}_2$ and ${\cal H}_3$ should enhance
antiferromagnetic correlations,
the ferromagnetically ordered state becomes unstable.
We find that a zigzag magnetically ordered state is realized
with finite $m^z_S$ around $K/U\sim 0.12$.
To study the competition between these ordered states, we show
the ground state energies in the inset of Fig.~\ref{fig:mag}.
We clearly find the hysteresis in the curves,
which indicates the existence of the first-order phase transition.
By examining the crossing point,
we clarify that the quantum phase transition
between ferromagnetically and zigzag ordered states occurs at $K/U\sim 0.15$.
In the case with $K/U<0.1$, due to strong frustration,
it is hard to obtain the converged solutions.
This will be interesting to clarify this point in a future investigation.

\begin{figure}
\begin{center}
\includegraphics[width=6cm,angle=-90]{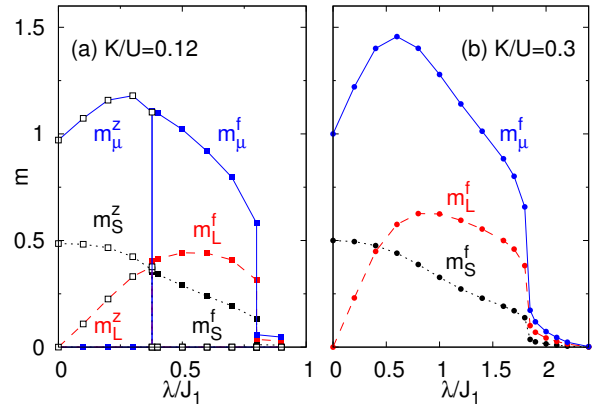}
\end{center}
\caption{
Total magnetic moment $m_\mu$, spin moment $m_S$, and orbital moment $m_L$
in the spin-orbital systems with (a) $K/U=0.12$ 
 and (b) $K/U=0.3$. 
}
\label{fig:mag3}
\end{figure}

The introduction of $\lambda$ couples the spin and orbital degrees of freedom.
The spin moments slightly decrease in both states,
as shown in Fig. \ref{fig:mag}.
The zigzag and ferromagnetically ordered states are stable against
the small SO coupling and
the first-order transition point has little effect on the SO coupling.
To discuss the stability of these states against the strong SO coupling,
we calculate the spin and orbital moments
in the system with $K/U=0.12$ and $0.3$, as shown in Fig.~\ref{fig:mag3}.
The introduction of the SO coupling slightly decreases
the spin moment, as discussed above.
By contrast, the orbital moment is induced parallel to the spin moment.
Therefore, the total magnetic moment $m_\mu^\alpha=|\sum_i(-1)^{\delta_i^\alpha}
\langle 2{\bf S}_i + {\bf L}_i \rangle|/N$ increases.
When $K/U=0.12$, the zigzag ordered state becomes unstable and
the first-order phase transition occurs
to the ferromagnetically ordered state at $\lambda/J_1\sim 0.4$.
Further increase of the SO coupling decreases the total moment $m_\mu^f$.
Finally, a jump singularity appears around $\lambda/J_1\sim 0.8 (1.8)$
in the system with $K/U=0.12 (0.3)$.
It is also found that
the magnetic moment is almost zero and each orbital is equally occupied
as in the isospin states $|\tilde{\sigma}\rangle$
in the larger SO coupling region.
Therefore, we believe that this state is 
essentially the same as the QSL state realized in the Kitaev model.

\begin{figure}
\begin{center}
\includegraphics[width=7cm]{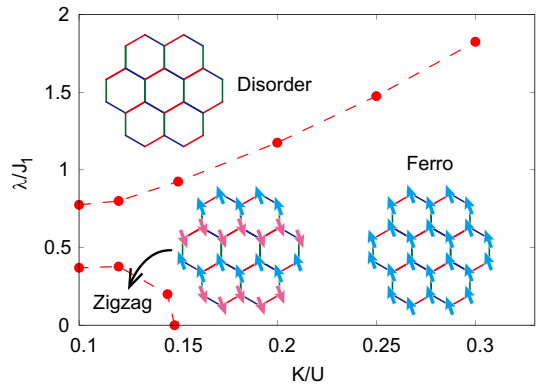}
\end{center}
\caption{
The ground state phase diagram of the spin-orbital model.
Transition points are obtained by the CMF-10.
}
\label{fig:gspd}
\end{figure}

By performing similar calculations,
we obtain the ground state phase diagram, as shown in Fig.~\ref{fig:gspd}.
The disordered (QSL) state is realized in the region with
large $\lambda/J_1$.
The ferromagnetically ordered state is realized
in the region with small $\lambda/J_1$ and large $K/U$.
The decrease of the Hund coupling induces spin frustration, which destabilizes
the ferromagnetically ordered state.
We wish to note that the zigzag ordered state is stable
in the small SO coupling region,
which is not directly taken into account in the Kitaev model.


Next, we discuss thermodynamic properties in the system.
It is known that, in the Kitaev limit ($\lambda\rightarrow\infty$),
the excitations are characterized by two energy scales, which correspond
to localized and itinerant Majorana fermions.
This clearly appears in the specific heat as two peaks
at $T/{\tilde J}=0.012$ and $0.38$~\cite{Nasu2015}.
To clarify how the double peak structure appears
in the intermediate SO coupling region,
we make use of the TPQ state for the twelve-site
cluster with the periodic boundary condition~[see Fig. \ref{fig:lattice}(b)].
According to the previous study~\cite{YamajiTPQ},
the double peak structure appears in the spin-1/2 
 Kitaev model
even with the twelve-site cluster.
Therefore, we believe that thermodynamic properties in the system
can be discussed, at least, qualitatively in our calculations.

\begin{figure}
\begin{center}
\includegraphics[width=7.5cm]{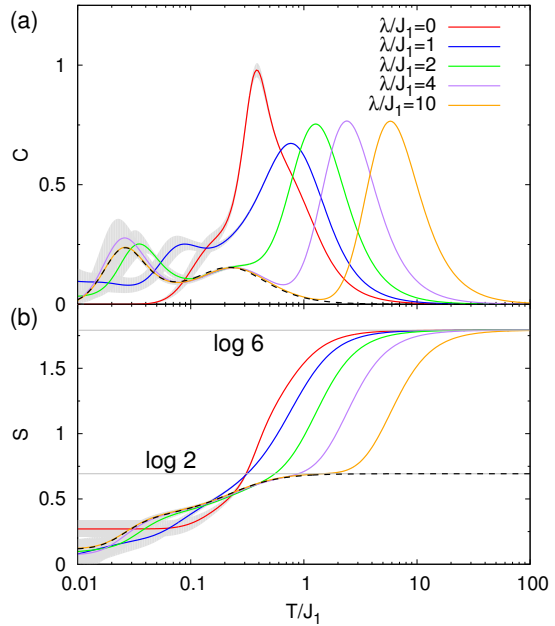}
\end{center}
\caption{
The specific heat (a) and entropy (b) as a function of the temperature
for the system with $\lambda/J_1=0, 1, 2, 4$, and $10$.
Dashed lines represent the results for the isospin Kitaev model
with twelve sites.
}
\label{fig:tpq}
\end{figure}

Here, we fix the Hund coupling as $K/U=0.3$
to discuss finite temperature properties
in the system with the intermediate SO coupling.
Figure~\ref{fig:tpq} shows the specific heat and
entropy in the system with $\lambda/J_1=0, 1, 2, 4$ and $10$.
In this calculation, the quantities are deduced by
the statistical average of the results
obtained from, at least, twenty independent TPQ states.
When $\lambda=0$,
we find a broad peak around $T/J_1=0.4$ in the curve of the specific heat.
In addition, most of the entropy is released at $T/J_1\sim 0.1$,
as shown in Fig.~\ref{fig:tpq}(b).
This can be explained by the fact that ferromagnetic correlations are enhanced
and spin degrees of freedom are almost frozen.
The appearance of the large residual entropy should be an artifact
in the small cluster with the orbital frustration.
The introduction of the SO coupling leads to interesting behavior.
It is clearly found that the broad peak shifts to higher temperatures.
This indicates the formation of the Kramers doublet
and a part of the entropy $S=\log(6)-\log(2)$ is almost released,
as shown in Fig.~\ref{fig:tpq}(b).
In addition, we find in the case $\lambda/J_1\ge 2$,
two peaks in the specific heat at lower temperatures.
The corresponding temperatures are little changed by the magnitude of
the SO coupling and the curves are quantitatively consistent with
the results for the isospin Kitaev model
on the twelve sites, which are shown as dashed lines.
Therefore, we believe that the Kitaev physics appears in the region.
On the other hand, when $\lambda/J_1=1$,
a single peak structure appears in the specific heat,
indicating that the Kitaev physics is hidden
by the formation of the Kramers doublet due to the competition
between the exchange interaction and SO coupling.
We have used the TPQ states to clarify how the double peak structure
inherent in the Kitaev physics appears,
in addition to the broad peak for the formation of the Kramers
doublet at higher temperatures.

To conclude, we have studied the effective spin-orbital model
obtained by the second-order perturbation theory.
Combining the CMF theory with
the ED method,
we have treated the Kugel-Khomskii type superexchange interaction and
SO coupling on an equal footing
to determine the ground-state phase diagram.
We have clarified how the magnetically ordered state competes with
the nonmagnetic state, which is adiabatically connected to
the QSL state realized in a strong SO coupling limit.
Particularly, we have revealed that a zigzag ordered state is realized
in this effective spin-orbital model with finite SO couplings.
The present study suggests another mechanism to stabilize the zigzag ordered phase close to the QSL in the plausible situation, and also will stimulate further experimental studies in the viewpoint of the SO coupling effect on magnetic properties in Kitaev candidate materials.

\begin{acknowledgments}
The authors would like to thank S. Suga for valuable discussions.
Parts of the numerical calculations are performed
in the supercomputing systems in ISSP, the University of Tokyo.
This work was partly supported by the Grant-in-Aid for
Scientific Research from JSPS, KAKENHI Grant Number
JP17K05536, JP16H01066 (A.K.), and JP16K17747, JP16H00987 (J.N.).
\end{acknowledgments}

\end{document}